\documentclass{article}
\usepackage{spconf,amsmath,graphicx,hyperref}
\hypersetup{hidelinks}

\usepackage{balance}
\usepackage{cite}
\usepackage{amsmath}
\DeclareMathOperator*{\med}{med}
\usepackage{amssymb}
\usepackage{amsfonts}
\usepackage{graphicx}
\usepackage{textcomp}
\usepackage{xcolor}
\usepackage{siunitx}
\usepackage{booktabs}
\usepackage{caption}
\usepackage{subcaption}
\usepackage{cleveref}
\crefname{figure}{Fig.}{Figs.}
\Crefname{figure}{Fig.}{Figs.}
\crefname{table}{Table}{Tables}
\Crefname{table}{Table}{Tables}
\usepackage[section]{placeins}
\usepackage{float}

\graphicspath{{figures/}{figures/synthetic/}{figures/real/}}

\newcommand{\cyc}{\alpha}

\title{Bio-inspired efficient cyclostationary analysis in machine and underwater acoustic recordings}
\name{Thura Pyae Sone, Abishek Soti, Henry Zhong, Gregory Cohen, RunChun Mark Wang, Ying Xu}
\address{ICNS, Western Sydney University, Sydney, Australia}

\begin{document}

\maketitle

\begin{abstract}
We propose a bio-inspired approach that uses the inner-hair-cell (IHC) response of the Cascade of Asymmetric Resonators with Fast-Acting Compression (CARFAC) model to efficiently extract cyclic modulation from acoustic signals. We further investigate the contribution of IHC processing by comparing the CARFAC-IHC response with the CARFAC basilar-membrane (BM) filtering. Furthermore, the CARFAC-IHC and CARFAC-BM approach are benchmarked against conventional FFT Accumulation Method (FAM), Integrated Cyclic Modulation Coherence (ICMC), and Detection of Envelope Modulation On Noise (DEMON) approaches using the Case Western Reserve University (CWRU) bearing dataset and a real ShipsEar work-vessel recording dataset. The results demonstrate reliable recovery of characteristic cyclic components while substantially reducing the computational burden of conventional cyclostationary analysis.
\end{abstract}


\section{Introduction}

Passive listening to a vessel's radiated noise is attractive for underwater ship
detection, since a vessel's rotating and reciprocating machinery imprints a
cyclostationary modulation signature on its radiated noise
\cite{Gardner1986,Gardner1987,Gardner1991,Antoni2009,AntoniBonnardot2004}. In real
ship-radiated noise, however, this signature is often weak, distributed across broad
frequency bands, and further obscured by ambient noise and non-stationary interference
\cite{Ollivier2021}.

Conventionally, we use Detection of Envelope Modulation On Noise (DEMON) processing to
\cite{Pollara2016} extract envelope modulation via band-pass filtering and envelope
detection of a target, but this approach does not fully exploit the target's frequency-dependent structure. Therefore, a different approach, 
spectral-kurtosis-guided envelope analysis \cite{AntoniRandall2006}, is adapted to sharpen DEMON's
band selection which is a special case of the same underlying cyclostationary
theory \cite{AntoniHanson2012}. 
Alternatively, cyclostationary spectral methods are also used instead to compute a
full spectral correlation surface with the FFT Accumulation Method (FAM)
\cite{Gardner1991} as a front end, then convert that surface into a one-dimensional
detection profile through an integrated cyclic modulation coherence (ICMC)
normalisation step \cite{AntoniHanson2012,Tong2024,Ollivier2021}. However, FAM with ICMC has high computational complexity and is susceptible to artefact generation in the time-frequency decomposition.
 
To address those problems, we propose a bio-inspired CARFAC cochlear processing approach for cyclostationary
estimation.  The CARFAC filterbank decomposes the signal into
frequency-localised channels, and its IHC channels' output extract envelope modulation. 
It can be integrated across channels to produce cyclic signatures.  In this work, we investigate the feasibility of the CARFAC-IHC cyclostationary extraction and compare the performance with the FAM/ICMC
pipeline. 

\section{Method}
\label{sec:method}

\subsection{CARFAC cochlear processing}
\label{sec:carfac}

\begin{figure}[H]
\centering
\includegraphics[trim=25 245 25 245, clip, width=0.9\linewidth]{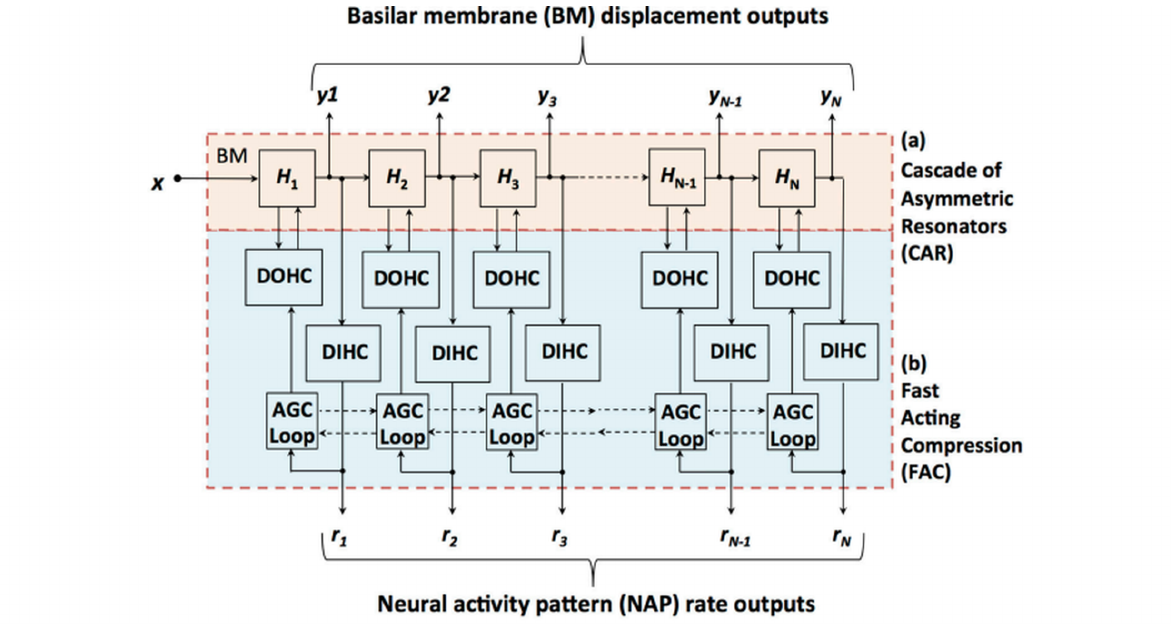}
\caption{CARFAC block diagram (adapted from \cite{Xu2018FPGA}): the CAR cascade
produces BM outputs $y_1,\dots,y_N$, with per-channel Automatic Gain Control feedback
setting the damping $r_1,\dots,r_N$.}
\label{fig:carfac-diagram}
\end{figure}

CARFAC processes the input waveform as 
shown in \cref{fig:carfac-diagram}. First, a cascade of asymmetric resonators (CAR) with centre frequencies $f_c$
placed by the Greenwood cochlear-position function~\cite{Greenwood1990} creates
a log-frequency decomposition. Second, a digital outer-hair-cell (DOHC) provides a feedback path to adjust the BM damping parameters via
an instantaneous feedback and an automatic-gain-control (AGC) feedback path. The AGC includes a four stage temporal smoothing filter spanning
a range of time constants with three spatial smoothing at each temporal stage.

Next, a digital inner-hair-cell (DIHC) detection stage (half-wave rectification followed by a
leaky integrator with multiplicative transmitter depletion~\cite{Lyon2011} extracts the modulation information carried by $y_s^{\mathrm{IHC}}[n]$ in each cochlear channel $s$.

\begin{table}[H]
\centering
\small
\caption{CARFAC envelope-spectrum estimator settings.}
\label{tab:carfac-params}
\setlength{\tabcolsep}{3pt}
\begin{tabular}{@{}lll@{}}
\toprule
Symbol & Meaning & Value \\
\midrule
$S$ & cochlear sections & $64$ \\
$[f_{\mathrm{lo}},f_{\mathrm{hi}}]$ & CAR passband & $\approx\!20$\,Hz--$0.45f_s$ \\
$\tau_{\mathrm{IHC}}$ & IHC leaky integrator & \SI{1.6}{\milli\second} ($\approx\!\SI{100}{\hertz}$) \\
$f_s'$ & envelope rate (downsampled) & $\max(512,\,2.5\,\cyc_{\max})$\,\si{\hertz} \\
$T_{\mathrm{win}}$ & envelope history & \SI{8}{\second} \\
$N_{\mathrm{F}}$ & envelope FFT length & $4096$ \\
--- & channel norm.\ / aggregation & local (\SI{\pm 20}{\hertz}) / median \\
\bottomrule
\end{tabular}
\end{table}

\subsection{CARFAC-IHC envelope spectrum}
\label{sec:carfac-alpha}

We extract cyclostationary signatures from the CARFAC-IHC channels: 
$\mathbf{y}^{\mathrm{IHC}}[n]$ (\cref{tab:carfac-params}) preserves the target signal modulation on the envelope. Downsampling is used for the cyclic frequencies of
interest ($\alpha \le \cyc_{\max}$), which are far below the IHC sampling rate.
The downsampling
computes the frame-based energy on each IHC channel as following:
\begin{equation}
  e_s[\ell] = \tfrac1D\!\sum_{i=0}^{D-1} y^{\mathrm{IHC}}_s[\ell D+i]^{2},
  \label{eq:pool}
\end{equation}

\noindent where in each channel $s$, $y^{\mathrm{IHC}}_s[n]$ is partitioned into non-overlapping blocks of $D$. The squared values within a block are then averaged to yield a single downsampled energy sample $e_s[\ell]$ per block, at the reduced rate $f_s'$.

While in CARFAC, the IHC higher frequency channels response tends to turn its frequency output into a DC offset. Therefore, we calculate $e_s[\ell]$ which is mean-subtracted and high-pass filtered
(\SI{5}{\hertz} cutoff) to remove its DC/near-DC offset.
The magnitude of the mean-removed energy envelope is calculated as follows.
\begin{equation}
  P_s(\cyc_j) = \Bigl|\ \sum_{\ell=0}^{N_{\mathrm{F}}-1} \bigl(e_s[\ell]-\bar e_s\bigr)\,
      e^{-\jmath 2\pi j\ell/N_{\mathrm{F}}}\ \Bigr|.
  \label{eq:chan-fft}
\end{equation}

where the per-channel energy trace from Eq. \eqref{eq:pool},  giving $P_s(\cyc_j)$  the strength of cyclic modulation at frequency $\cyc_j$ in that channel.

A single per-channel scalar normalisation (e.g.\ dividing by
$\max_{\cyc'}P_s(\cyc')$) leaves a weak, phase-locked line undetectable
whenever a stronger, unrelated line dominates that same channel, so we
normalise per alpha bin, dividing by the local median power in a
$\pm\Delta\cyc$ neighbourhood of that bin. Channels are then combined with a further median,
across $s$ this time rather than across $\cyc$, to reduce square-law self-noise inflating the aggregate.
\begin{equation}
  A(\cyc_j) = \med_{s=1,\dots,S}\ \frac{P_s(\cyc_j)}
      {\med\limits_{\cyc'\in[\cyc_j-\Delta\cyc,\,\cyc_j+\Delta\cyc]} P_s(\cyc')},
  \label{eq:Aalpha}
\end{equation}
with $\Delta\cyc=\SI{20}{\hertz}$, retained for $\cyc_j\in[\cyc_{\min},f_s'/2]$.
For visualisation, the plotted spectrum uses this same aggregation with
$\max_{\cyc'}P_s(\cyc')$. 

Unlike FAM, which
forms its spectral correlation surface by cross-multiplying frequency-shifted
spectra with their complex conjugate, $P_s(\cyc_j)$ in \eqref{eq:chan-fft} is
a plain magnitude spectrum computed from the real-valued energy signal $e_s[\ell]$:
there is no conjugate cross-term and no spectral correlation surface to compute.

\subsection{CARFAC-BM into FAM/ICMC}
\label{sec:carfac-bm}

Here the CAR passband is retuned from \cref{tab:carfac-params}'s wide setting to
the analysis band $[B_1,B_2]$ so that all $S$ Greenwood channels fall in band.
Retuning the span first guarantees every summed channel is on-band by construction,
avoiding channel reconstruction with a low-frequency cascade artefact.

Channels with the largest CARFAC resonance gain would otherwise dominate a
raw sum, so each channel is first scaled to unit RMS,
$\hat y^{\mathrm{BM}}_s = y^{\mathrm{BM}}_s\big/\mathrm{rms}(y^{\mathrm{BM}}_s)$,
before summing:
\begin{equation}
  x_{\mathrm{rec}}[n] = \sum_{s=1}^{S} \hat y^{\mathrm{BM}}_s[n].
  \label{eq:bm-recon}
\end{equation}

Equation \ref{eq:bm-recon} removes the relative amplitude information between channels, which is important for
lessening the effect of the CARFAC cascade's centre-frequency-dependent gain on the summed output. 

Since the CAR stage is a cascade, each channel's response also reaches the
output after a different, centre-frequency-dependent delay $\tau_s$. The delay calibration is
currently omitted because it is a linear, small-signal measurement, while CARFAC's
actual operating delay is level-dependent through its AGC feedback and
amplitude-dependent damping, since a fixed calibration cannot track a delay that
shifts with the input's own signal level. The CARFAC-BM pipeline
therefore keeps only the amplitude fix. The result is then mean-removed, RMS-normalised, and decimated to
$f_s^{\mathrm{ICMC}}=\min(f_s^{\mathrm{in}},f_s^{\mathrm{CAR}})$, then passed
 to the standard FAM $\rightarrow$ spectral-coherence
$\rightarrow$ band-integrated cyclic-modulation-profile pipeline (ICMC).

\section{Results: Synthetic Data}
\label{sec:results-synthetic}

Firstly, we use a full-carrier double-sideband amplitude-modulated (DSB-AM) signal to test the proposed CARFAC-IHC approach against the FAM/ICMC baseline and the CARFAC-BM on known analytical results. \cite{Gardner1987}
\begin{equation}
  x(t) = \left[1 + m\cos(2\pi f_m t)\right]\cos(2\pi f_c t),
  \label{eq:dsbam}
\end{equation}
with sampling rate $f_s=\SI{32000}{\hertz}$, carrier frequency
$f_c=\SI{1000}{\hertz}$, modulation frequency $f_m=\SI{100}{\hertz}$, and
modulation index $m=0.8$. From the standard worked example in cyclostationary theory \cite{Gardner1987}, 
the mean power is at $\cyc=0$, with envelope modulation frequencies $\cyc=\pm f_m$, and the carrier self-beating and its sidebands at $\cyc=\pm 2f_c,\ \pm2f_c\pm f_m$. 
\begin{table}[H]
\centering
\small
\caption{FAM/ICMC channeliser parameters.}
\label{tab:sim-parameters}
\setlength{\tabcolsep}{4pt}
\begin{tabular*}{\linewidth}{@{\extracolsep{\fill}}llc@{}}
\toprule
Symbol & Description & Value \\
\midrule
$N_p$   & FAM channeliser FFT length      & 1024 \\
$\Delta f$ & FAM frequency resolution ($f_s/N_p$) & \SI{31.25}{\hertz} \\
$L$     & FAM frame hop size   & 256 ($N_p/4$) \\
$p$   & FAM number of frames       & 497 \\
\bottomrule
\end{tabular*}
\end{table}

\begin{table}[H]
\centering
\small
\setlength{\tabcolsep}{3pt}
\renewcommand{\arraystretch}{0.9}
\caption{Estimated vs.\ theoretical cyclic frequencies for the synthetic double-sideband AM tone of \eqref{eq:dsbam}, recovered by CARFAC-IHC (\cref{sec:carfac-alpha}) on the same signal used to validate CARFAC-BM.}
\label{tab:synthetic-validation-ihc}
\begin{tabular}{@{}lrrr@{}}
\toprule
Signal feature & Predicted (Hz) & Estimated (Hz) & Error (Hz) \\
\midrule
$|\alpha| = f_m$ (envelope) & 100.000 & 99.935 & $-$0.065 \\
$|\alpha| = 2f_c$ & 2000.000 & 2000.000 & +0.000 \\
$|\alpha| = 2f_c + f_m$ & 2100.000 & 2099.935 & $-$0.065 \\
$|\alpha| = 2f_c - f_m$ & 1900.000 & 1900.065 & +0.065 \\
\bottomrule
\end{tabular}
\par\smallskip\small Errors are at most 0.065~Hz, under half the resolution bin ($\Delta\alpha = 0.163$~Hz).
\end{table}

\Cref{tab:synthetic-validation-ihc} validates CARFAC-IHC (\cref{sec:carfac-alpha})
on the identical signal. Each CARFAC-IHC channel is rectified independently, with no
cross-channel coherence step. The true signal is a three-tone sum ($f_c-f_m$, $f_c$,
$f_c+f_m$), and squaring introduces a sum or difference term for every pair
of tones a channel responds to, not only the pairs Gardner's simplified
treatment highlights.

\begin{table}[H]
\centering
\small
\setlength{\tabcolsep}{3pt}
\renewcommand{\arraystretch}{0.9}
\caption{Estimated vs.\ theoretical cyclic frequencies for the synthetic double-sideband AM tone of \eqref{eq:dsbam}, recovered by the CARFAC-BM validation route.}
\label{tab:synthetic-validation-carfac}
\begin{tabular}{@{}lrrr@{}}
\toprule
Signal feature & Predicted (Hz) & Estimated (Hz) & Error (Hz) \\
\midrule
$\alpha = 0$ (mean power) & 0.000 & -0.063 & -0.063 \\
$|\alpha| = f_m$ (envelope) & 100.000 & 100.038 & +0.038 \\
$|\alpha| = 2f_c$ & 2000.000 & 2000.000 & +0.000 \\
$|\alpha| = 2f_c + f_m$ & 2100.000 & 2100.038 & +0.038 \\
$|\alpha| = 2f_c - f_m$ & 1900.000 & 1899.962 & -0.038 \\
\bottomrule
\end{tabular}
\par\smallskip\small Errors are at most 0.063~Hz, all under half the cyclic-frequency resolution bin ($\Delta\alpha = 0.2515$~Hz).
\end{table}

\Cref{tab:synthetic-validation-carfac} validates the CARFAC-BM route
on this signal. All theoretical cyclic frequencies are recovered. The cochlear front end therefore
preserves the second-order cyclostationary structure the estimator relies on.



\section{Results: Real Data}
\label{sec:results-real}

Next, we test the approaches on  both Case Western Reserve University (CWRU) and ShipsEar recordings.

\subsection{CWRU bearing dataset}

\begin{figure}[H]
\centering
\includegraphics[width=\linewidth]{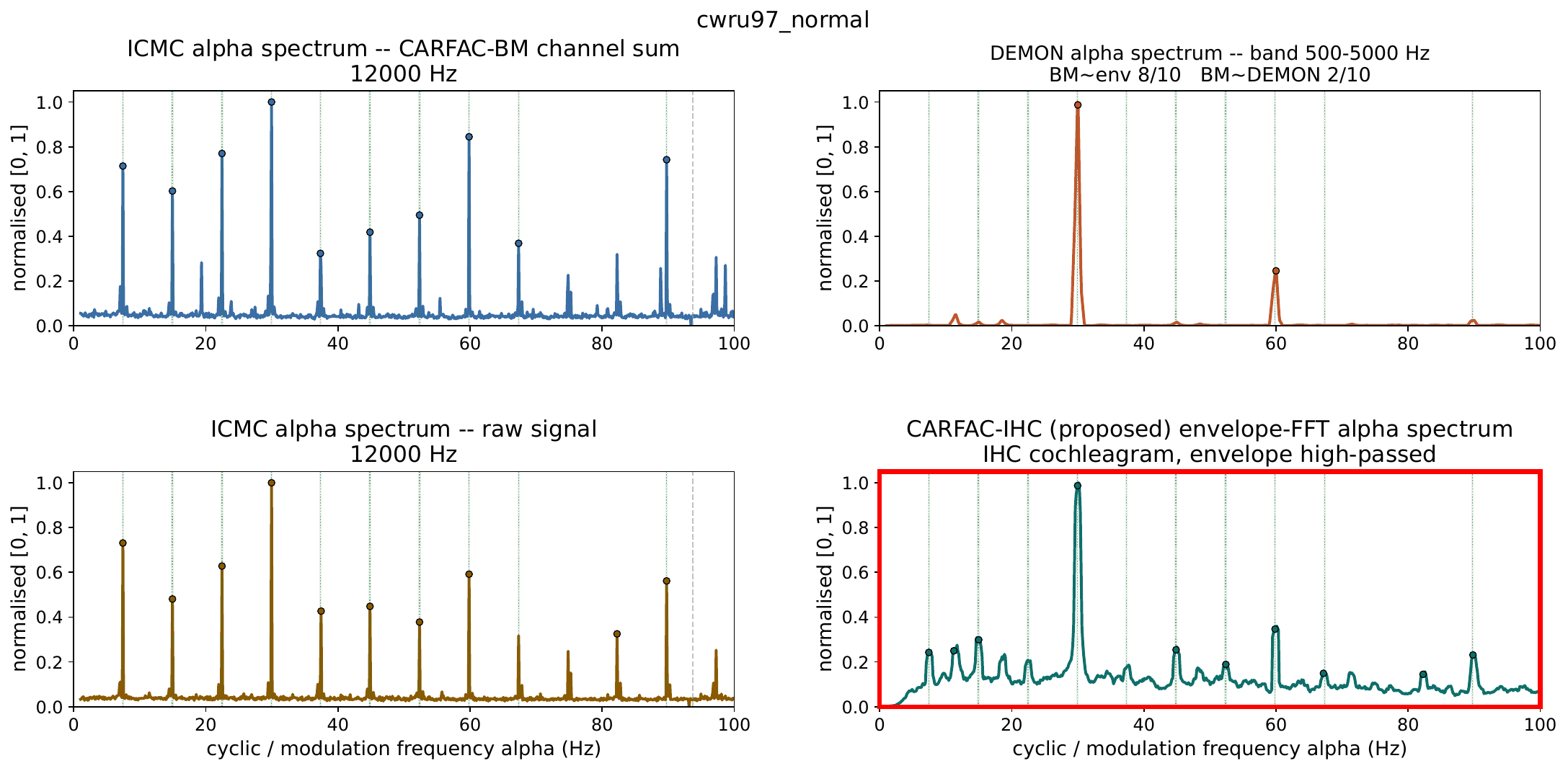}
\caption{Cross-estimator validation on a healthy CWRU Bearing
Data Center record (0\,hp load, \SI{1797}{rpm}): ICMC on the CARFAC-BM
reconstruction, DEMON, raw-signal ICMC, and CARFAC-IHC alpha spectra.}
\label{fig:cwru-carfac-frontend}
\end{figure}

The CWRU Bearing Data Center dataset
\cite{CWRUBearingDataCenter,SmithRandall2015} is used here on a healthy,
unfaulted bearing (0\,hp load, \SI{1797}{rpm}). As shown in \cref{fig:cwru-carfac-frontend}, the proposed CARFAC-IHC (bottom right) extracts both the
shaft rotation frequency $f_r=\mathrm{rpm}/60\approx\SI{30}{\hertz}$ and its
harmonics $kf_r$ (i.e., 30, 60, and 90\,Hz), present in the signal. The DEMON approach (top right) extracts only 30 and 60\,Hz. Both CARFAC-BM (top left) and ICMC (bottom left) show 30\,Hz and its harmonics. The recording also contains a quarter-order harmonic series: CARFAC-BM and ICMC additionally
resolve a 7.5\,Hz modulation and its overtones, and CARFAC-IHC resolves the same
7.5\,Hz component, though noisily. DEMON does not recover this modulation.

\subsection{ShipsEar dataset}

\begin{figure}[H]
\centering
\includegraphics[width=1\linewidth]{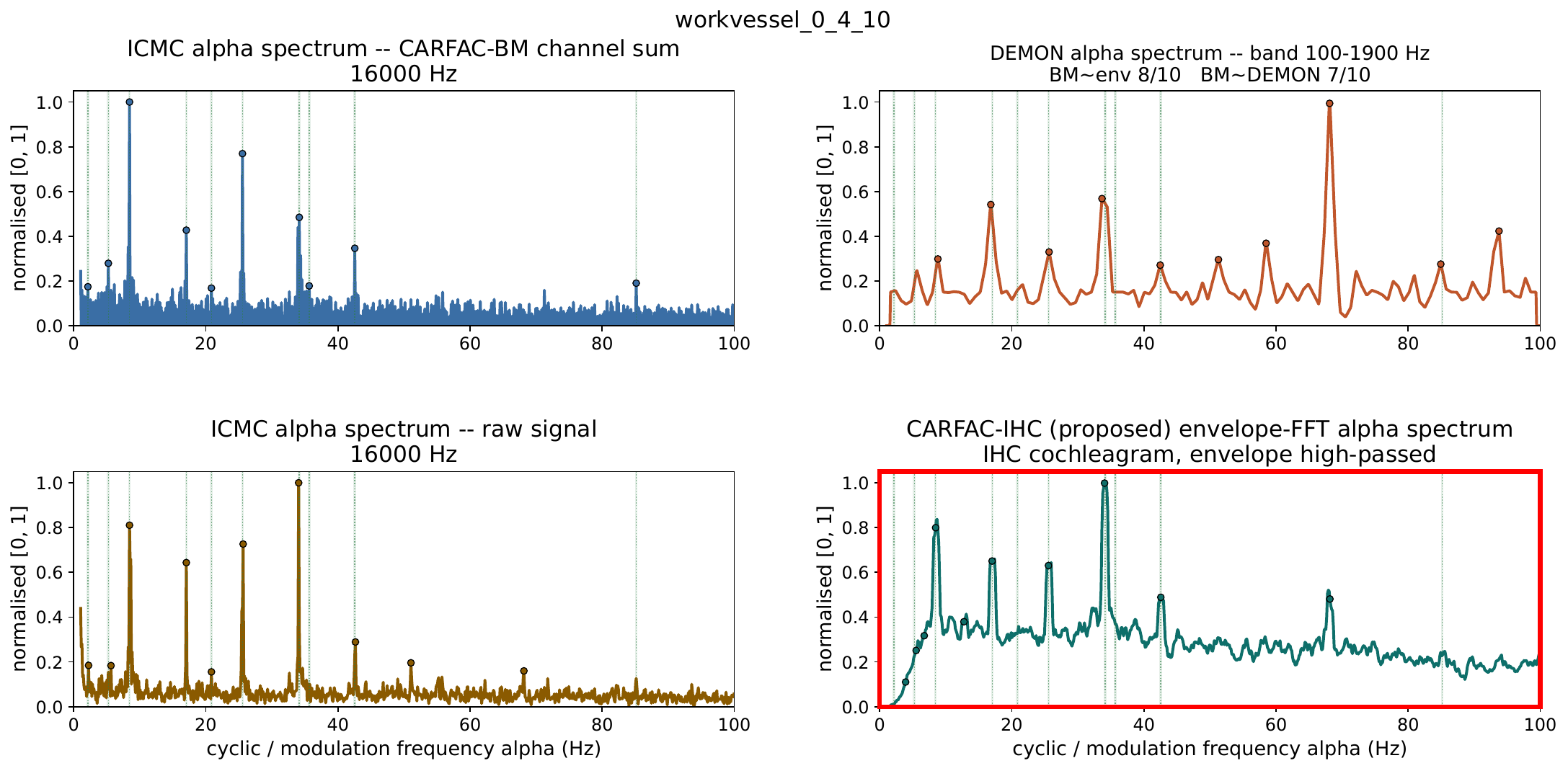}
\caption{Cross-estimator validation on a work-vessel recording from the public
ShipsEar dataset \cite{SantosDominguez2016}, processed at \SI{16}{\kilo\hertz} over the
\SIrange{100}{1900}{\hertz} band.}
\label{fig:workvessel-carfac-frontend}
\end{figure}

ShipsEar \cite{SantosDominguez2016} provides a reproducible, public benchmark for
the cross-estimator pipeline, processed at \SI{16}{\kilo\hertz} over the
\SIrange{100}{1900}{\hertz} band. Unlike CWRU, ShipsEar carries no documented shaft speed or gear geometry, so there is
no independent theoretical line to check against. \cref{fig:workvessel-carfac-frontend} shows the result of cyclostationary analysis with the same peak-detection
standard as CWRU, counting only distinct confirming peaks, the CARFAC-BM variant
tracks raw ICMC on 9 of 10 compared peaks, agrees with the DEMON on 7 of 10, and with
the CARFAC-IHC on 8 of 10. The empirically checked CARFAC-IHC results has shown that cyclostationary is also viable in underwater acoustics.

Additionally, the CARFAC-IHC result has a visibly higher noise floor on a real-world hydrophone recording since it carries broadband ambient/flow noise
shared across most of the 64 channels rather than isolated to a few, so the
per-channel and cross-channel median normalisation of \cref{sec:carfac-alpha} produce a comparatively poorer result to CWRU dataset.

\section{Discussion}

\subsection{Computational considerations and generalisability}
\label{sec:computational}
\begin{figure}[H]
\centering
\includegraphics[width=\linewidth]{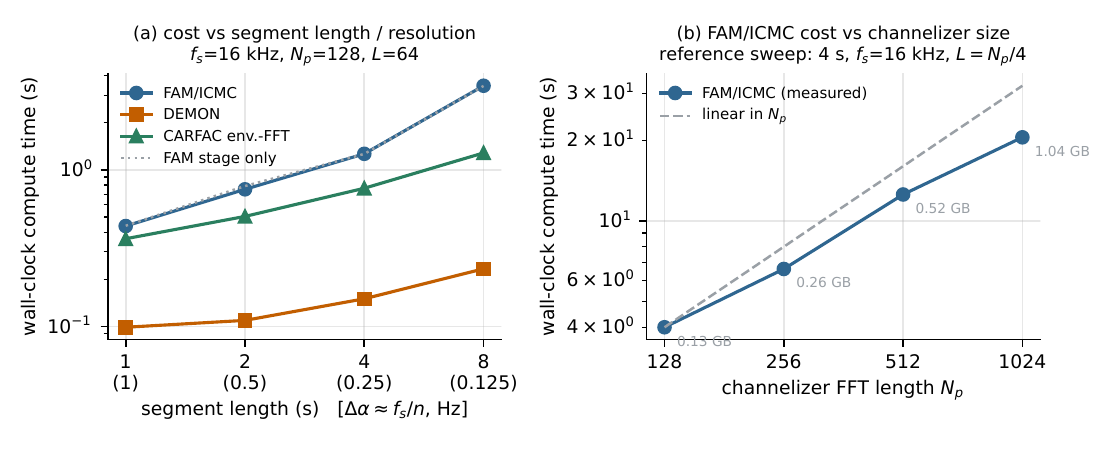}
\caption{Wall-clock compute time versus (a) segment length/resolution at
$f_s=\SI{16}{\kilo\hertz}$ and (b) channeliser size $N_p$ for FAM/ICMC, with
accumulator memory annotated. CARFAC-IHC and DEMON scale far
more gently than FAM/ICMC, consistent with a real-time, edge-deployable implementation.}
\label{fig:compute-scaling}
\end{figure}

Edge-device deployment imposes strict computational budgets, and FAM/ICMC is the
most expensive of the three estimators: it is an inherently batch computation
backed by a memory upper bound of $O(P N_p^{2})$ in the accumulator that exceeds \SI{1}{\giga\byte} at
$N_p=1024$, and this cost increases with the channeliser size $N_p$
(\cref{fig:compute-scaling}). The CARFAC-BM
validation route inherits this same cost by design since it prepends a per-sample
cochlear pass to the FAM/ICMC back end, adding cochleagram cost on top.

CARFAC-IHC is a different estimator off the same front end. Like DEMON, it dispenses with the block sweep entirely, and both scale far more gently
with segment length and resolution than FAM/ICMC does, as shown in \cref{fig:compute-scaling}.
Measured directly against FAM/ICMC's own wall-clock cost on the same task, DEMON
and CARFAC-IHC take \numrange{13}{102}$\times$ and \numrange{2}{37}$\times$ less
compute time respectively, and the gap widens further as $N_p$ increases. These
figures were measured single-threaded on a desktop-class CPU, substantially more
capable than the embedded processors typical of underwater edge platforms, so
FAM/ICMC's cost disadvantage would be more pronounced under actual deployment
hardware. A deployed system can
therefore run this cheap route continuously and call FAM/ICMC only to confirm a
candidate. The CARFAC implementations are already FPGA-proven for underwater sensing
\cite{Xu2018FPGA,Bremer2025}, so the streaming route this paper proposes is
edge-deployable as it stands.

The front end's usefulness is also not confined to the ICMC/DEMON comparison made
here. Because CARFAC's cochleagram is a general per-channel envelope
representation, the same architecture could feed any downstream cyclostationary
detector, extending beyond the propeller and bearing-fault applications
demonstrated in this paper. All wall-clock timings reported above were measured
on a single machine (Intel Core i7-9700K CPU @ \SI{3.60}{\giga\hertz}, 8 cores,
\SI{32}{\giga\byte} RAM), single-threaded.

The principal limitation of the conventional FAM estimator is structural. As the periodic block advance, $L$ injects a signal-independent spurious correlation at $\alpha = k f_s/L$ and $\alpha = k f_s/N_p$ ($k=1,2,\dots$), DEMON and our proposed CARFAC-IHC estimator (\cref{sec:carfac-alpha}), neither tied to $L$ or $N_p$ since they are not block-based.

An extension to the project is to produce a
fully bio-inspired auditory representation: CAR-FAC paired with 
neuron models \cite{Xu2023EventDriven} to build an auditory sensing pathway, and
detecting cyclostationary structure directly in that stream would replace this
paper's fixed analysis windows with an asynchronous, transient-triggered one, cutting
the power budget by orders of magnitude relative to \cref{sec:computational}.

Beyond detection, the same front end could close the loop into autonomous
control: sound-source localisation already informs autonomous robot
navigation \cite{Jalayer2025}, and FPGA-based edge perception is already
deployed on Autonomous underwater vehicles (AUVs)\cite{Xu2018FPGA,Bremer2025,Qi2025,Zhao2019}, making an AUV homing on a detected signature a natural
extension.

\subsection{Conclusion}
We propose the use of CARFAC-IHC output in cyclostationary analysis, turning an
existing, well-studied cochlear model into a practical cyclic-modulation
detector. Across synthetic, CWRU, and
ShipsEar recordings, the CARFAC-IHC recovers the majority of the low frequency
cyclic structure up to $37\times$ faster than FAM/ICMC
(\cref{sec:computational}). This margin makes
CARFAC-IHC the practical choice for continuous monitoring, with a clear path
toward the autonomous, closed-loop applications.

\clearpage
\balance

\bibliographystyle{IEEEbib}
\bibliography{references}

\end{document}